\documentclass[twocolumn, tighten]{aastex701}
\usepackage{mathtools}
\usepackage{orcidlink}

\begin{document}

\title{The role of Bethe-Heitler pair production in reconnection-driven flares in M87*}

\author[orcid=0000-0001-6640-0179,sname='Petropoulou']{Maria Petropoulou}
\affiliation{Department of Physics, National and Kapodistrian University of Athens, University Campus Zografos, GR~15784, Athens, Greece}
\affiliation{ Institute of Accelerating Systems \& Applications, University Campus Zografos, GR 15784, Athens, Greece}
\email[show]{mpetropo@phys.uoa.gr}  

\author[orcid=0000-0002-2383-3860, sname='Karavola']{Despina Karavola} 
\affiliation{Department of Physics, National and Kapodistrian University of Athens, University Campus Zografos, GR~15784, Athens, Greece}
\email[show]{dkaravola@phys.uoa.gr}

\author[orcid=0000-0002-1227-2754, sname='Sironi']{Lorenzo Sironi}
\affiliation{Department of Astronomy and Columbia Astrophysics Laboratory, Columbia University, New York, NY 10027, USA}
\affiliation{Center for Computational Astrophysics, Flatiron Institute, New York, NY 10010, USA}
\email{lsironi@astro.columbia.edu}

\begin{abstract}
Rapid TeV flares have been observed from the core of the active galaxy M87. These have been attributed to inverse Compton scattering of disk photons by electrons and positrons accelerated in transient reconnection layers formed in baryon-poor regions of the magnetosphere of the central black hole, M87*. It was previously shown that even a small number of protons accelerated in the same layers can lead to bright GeV proton-synchrotron flares, if protons receive $\gtrsim 20\%$ of the dissipated power for reconnecting fields of $\sim 100$ G. We aim to investigate the role of Bethe-Heitler pair production in the emission of reconnection-driven flares from M87* in this physical regime. We perform numerical calculations that incorporate inelastic collisions between relativistic protons and photons, as well as photon–photon pair production, and compute the non-thermal radiation from the layer. The numerical calculations are also supported by analytical estimates. We find that disk photons act as targets for Bethe–Heitler pair production. The resulting pairs emit very high-energy synchrotron photons ($\gtrsim 0.1$ TeV), which are subsequently attenuated by the disk photon field, leading to further pair production. The synchrotron emission of these secondary pairs produces soft photons, as part of an electromagnetic cascade, enhancing pion production and photon–photon attenuation down to tens-of-GeV energies. 
\end{abstract}


\section{Introduction}
Relativistic magnetic reconnection is a physical process through which magnetic fields of opposite polarity annihilate, rapidly converting magnetic energy into plasma kinetic energy. A substantial fraction of the dissipated energy is transferred to non-thermal particles accelerated by strong electric fields in the reconnection region on short timescales \citep[see][for a recent review]{2025ARA&A..63..127S}. Due to this combination of fast and efficient particle energization, relativistic reconnection has been widely invoked as a mechanism for powering rapidly varying high-energy flares in astrophysical systems, including blazars and gamma-ray bursts \citep[e.g.,][]{2009MNRAS.395L..29G, 2013MNRAS.431..355G, 2016MNRAS.455L...6B, 2016MNRAS.462.3325P}. 

M87 is a massive elliptical radio galaxy at a distance of $16.8$~Mpc and has a black hole mass of $6.5\cdot 10^9 M_{\odot}$ \citep{2019ApJ...875L...6E}. The non-thermal electromagnetic emission emerging from the vicinity of its supermassive black hole, M87*, is variable. For example, intraday X-ray variability from the core of M87 is occasionally observed \citep[e.g.,][]{2021ApJ...919..110I, 2023RAA....23f5018C}, as well as bright flares in TeV $\gamma$-rays on timescales as short as a few times the light crossing time of the black hole's gravitational radius \citep{2006Sci...314.1424A, 2009Sci...325..444A, 2012ApJ...746..151A}. These results hint to charged particle acceleration in the immediate vicinity of the black hole. Relativistic magnetic reconnection has been proposed as a promising explanation for the rapid activity from M87* and for accelerating particles to very high energies \citep[e.g.,][]{2023ApJ...943L..29H, 2024JCAP...12..009S, 2025ApJ...995L..73H}. 

In particular, \cite{2023ApJ...943L..29H} proposed that pairs accelerated by magnetic reconnection can inverse-Compton scatter low-energy disk photons to TeV energies, while releasing most of their energy via synchrotron radiation near the so-called burnoff limit, at a few tens of MeV energies. \cite{2025ApJ...995L..73H} (hereafter \href{https://ui.adsabs.harvard.edu/abs/2025ApJ...995L..73H}{H25}) extended this framework by considering the presence of protons in the reconnection region. They showed that even a minor proton population can significantly modify the predicted leptonic spectrum by producing a dominant component that peaks at $\sim10$ GeV through proton synchrotron radiation.

The aim of this Letter is to examine whether additional physical processes expected to operate in magnetospheric reconnection layers -- most notably Bethe–Heitler pair production -- can modify the key conclusions of  \href{https://ui.adsabs.harvard.edu/abs/2025ApJ...995L..73H}{H25}. In particular, we assess whether (i) $\sim 10$ GeV flares powered by proton synchrotron radiation can still carry a significant fraction of the magnetic energy dissipated during reconnection, and (ii) photopion interactions remain negligible once these additional channels are taken into account. We emphasize that this study is not intended as a full parametric exploration; instead, we demonstrate these effects for a representative set of parameters relevant to M87*. A more comprehensive investigation of the role of Bethe–Heitler and photon–photon processes for magnetospheric current sheets is deferred to future work.

To assess the importance of additional physical processes operating in the magnetospheric reconnection layer, we derive analytical expressions for the characteristic energies of the particles involved, as well as for the relevant optical depths. This analysis is complemented by radiative numerical calculations that incorporate inelastic collisions between relativistic protons and photons as well as photon–photon ($\gamma \gamma$) pair production, allowing the calculation of the escaping non-thermal radiation from the reconnection layer. The calculations, which are performed using the leptohadronic code ATHE$\nu$A \citep{2012A&A...546A.120D}, self-consistently include the feedback of secondary particles (both photons and pairs) produced in inelastic two-particle interactions on the cooling of relativistic protons and the attenuation of very high energy photons. We first demonstrate the impact of Bethe–Heitler and $\gamma \gamma$ pair production on the radiative output of the layer while neglecting the contribution of accelerated (primary) electron–positron pairs, thereby isolating the effects related to the hadronic channel. The role of primary pairs is examined later.

We find that soft photons from the accretion disk act as targets
for Bethe–Heitler pair production. The resulting Bethe–Heitler pairs emit very-high-energy (VHE, $\sim10$ GeV - 100 PeV) synchrotron photons, which are efficiently attenuated by the disk photon field ($\sim0.01$ eV), triggering further pair production. The synchrotron emission of these secondary pairs generates soft photons in the 0.1 - 100 eV range, enhancing photopion interactions and enabling photon–photon attenuation of gamma rays with energies as low as tens of GeV. As a consequence, the expected proton-synchrotron emission is strongly suppressed. Instead, a large fraction of power dissipated into protons is released in the form of high-energy neutrinos ($\sim100$ PeV) and reprocessed high-energy radiation. Finally, the network of processes initiated by the attenuation of Bethe-Heitler synchrotron photons produces a copious population of secondary pairs, which can even dominate the pair content of the plasma.

\section{Model}\label{sec:model}
We consider a current layer that forms in the magnetospheric region of the accreting black hole \citep{ripperda_black_2022}. We describe the current sheet as a cylindrical wedge of cross-sectional area $S = r^2/2$ (as in \href{https://ui.adsabs.harvard.edu/abs/2025ApJ...995L..73H}{H25}) and height $H=\beta_{\rm rec}r$,\footnote{This choice leads to a volume that is 4.5 larger than the one defined in \cite{2025ApJ...995L..73H} where the height is set by the Larmor radius of the most energetic protons in the system.} where $\beta_{\rm rec} \sim 0.1$ is the reconnection rate and $r \sim 10 \, R_{\rm g}$ is the radial extent of the current sheet \citep{ripperda_black_2022}. As the subsequent numerical calculations are tailored for a spherical geometry, we define an effective radius $R$ through the condition $4 \pi R^3/ 3 = SH$, resulting in $R=\left[3 \beta_{\rm rec}/(8 \pi) \right]^{1/3} r \sim 0.22 \, r$ \citep[see also][]{2024ApJ...961L..14F, karavola_neutrino_2025}. Mapping the cylindrical wedge to a sphere of equal volume preserves the particle number density and therefore ensures that two-particle interaction rates are correctly represented within the isotropic (angle-averaged) framework of the code ATHE$\nu$A. Alternative mappings that do not preserve particle density and depend on directionality are less appropriate for an angle-averaged treatment.

The plasma in the upstream region of the current sheet is expected to be pair dominated both in terms of mass and number (\href{https://ui.adsabs.harvard.edu/abs/2025ApJ...995L..73H}{H25}). Transient baryon-poor regions, extended over a large range of polar angles, are found to be formed in the magnetosphere of magnetically arrested disks (Chow et al., private communication). In this regime, the cold proton magnetization, $\sigma_{\rm p} = B^2/(4 \pi n_{\rm p} m_{\rm p} c^2) \gg 1$, is expected to be much larger than the total cold plasma magnetization, $\sigma = B^2/ [4 \pi (n_{\rm p} m_{\rm p} + n_{\pm} m_{\rm e})c^2]$. Here, $n_{\rm p}$ and $n_\pm$ denote the number density of protons and primary pairs, respectively. In this work, $\sigma_{\rm p}$ is a free parameter, but we discuss the impact of specific values in Appendix~\ref{appB}.

The inflow rate of electromagnetic energy through both sides of the current sheet is $L_{\rm EM} = c \beta_{\rm rec} B^2 (2S) / (4\pi) = c \beta_{\rm rec} B^2 r^2 / (4 \pi)$, where $B$ is the strength of the magnetic field in the upstream region of the current sheet. We adopt $B=100$~G as 
a representative value for the magnetosphere of M87*, in case the accretion is in the magnetically arrested state \citep[e.g.,][]{2021MNRAS.507.4864Y, 2023ApJ...944..173C, 2024JCAP...12..009S, 2025ApJ...995L..73H}. We demonstrate the effects of varying $B$ in Appendix~\ref{appB}. We assume that half of the Poynting luminosity is dissipated in the current sheet, $L_{\rm rec} = L_{\rm EM}/2$, and that a sizable fraction of it, ${\eta_{\rm p} \sim 0.1-0.2}$, is transferred to relativistic protons, $L_{\rm p} = \eta_{\rm p}L_{\rm rec}$. Such fractions are motivated by PIC simulations of reconnection in plasmas with mixed (i.e., pair-proton) compositions \citep{2019ApJ...880...37P, 2025ApJ...995L..73H}. Relativistic electrons and positrons that are accelerated in the current sheet (henceforth, primary pairs) receive the remaining power, $L_{\rm e} = (1-\eta_{\rm p}) L_{\rm rec}$. In the baseline model, we adopted $\eta_{\rm p} = 0.5$ as an optimistic upper-end value, chosen to highlight the regime in which proton-synchrotron emission is maximally relevant and consistent with the GeV predictions of \href{https://ui.adsabs.harvard.edu/abs/2025ApJ...995L..73H}{H25}, shown as the magenta shaded region in Fig.~\ref{fig:sed-1}. We explore lower values of $\eta_{\rm p}$ in Appendix~\ref{appB}.

The distribution function of accelerated protons is benchmarked against results from recent particle-in-cell (PIC) simulations of relativistic radiative reconnection in pair-proton plasmas (\href{https://ui.adsabs.harvard.edu/abs/2025ApJ...995L..73H}{H25}).  In particular, it is described by a broken power law, $dN_{\rm p}/d\gamma \propto \gamma^{-1}$ for $1 < \gamma \le \gamma_{\rm p, br}$ and $dN_{\rm p}/d\gamma \propto \gamma^{-s}$ for $\gamma_{\rm p, br} < \gamma \le \gamma_{\rm p, rad} $. The post-break slope $s$ depends on the strength of the guide field, as shown in \cite{com2024}; here, we adopt $s=2$, representative of weak guide fields, but we discuss other choices in Appendix~\ref{appB}. The simulations of \href{https://ui.adsabs.harvard.edu/abs/2025ApJ...995L..73H}{H25} \, further show that the break Lorentz factor is a fraction of the proton magnetization. Accordingly, we define $\gamma_{\rm p, br} = \xi \sigma_{\rm p}$ with $\xi = 1/3$ \footnote{The specific value of $\xi$ is not crucial since $\sigma_{\rm p}$ is a free parameter.}. The maximum Lorentz factor of the proton distribution is set by synchrotron radiative losses, $\gamma_{\rm p, rad} = (6 \pi e \beta_{\rm rec}/ (\sigma_{\rm T} B))^{1/2} (m_{\rm p}/m_{\rm e}) \simeq 10^{9.8} {(\beta_{\rm rec,-1})^{1/2} (B_2)^{-1/2}}$ (here, we introduced the notation $q_X = q/10^X$). This value is comparable to the Lorentz factor attained through acceleration in a reconnection electric field $\mathcal{E}_{\rm rec} = \beta_{\rm rec} B$ over the characteristic proton advection timescale of the layer, $R/c$, for the adopted values of $B$ and $R$ (see Table~\ref{params} and Fig.~7 in \cite{2024JCAP...12..009S}).

During the formation of the current sheet, accretion is temporarily halted, and the disk recedes to a distance comparable to the current sheet radius \citep{ripperda_black_2022}. If $L_{\rm ext}$ denotes the disk luminosity, its energy density within a spherical volume of radius $R$ is $u_{\rm ext} = f_{\rm d} L_{\rm ext}/(4 \pi R^2 c)$, where the factor $f_{\rm d}$ accounts for geometrical suppression of the flux. In the reconnecting region, the magnetic energy density typically exceeds that of the external radiation field. As a result, external photons primarily act as targets for secondary pair production, while synchrotron radiation remains the dominant emission and cooling mechanism of the resulting pairs. To illustrate this more clearly, we derive the magnetic field strength needed to achieve energy equipartition, $u_{\rm ext} = u_B$. This yields $B_{\rm eq} \simeq 2.7~{\rm G}~(R/2\,R_{\rm g})^{-4/3} {(\beta_{\rm rec, -1})^{-1/3}} f_{\rm d,0} \, L_{\rm ext, 42}$ which is much smaller than the adopted value of $B$.

\begin{figure}
    \centering
    \includegraphics[width=0.99\linewidth]{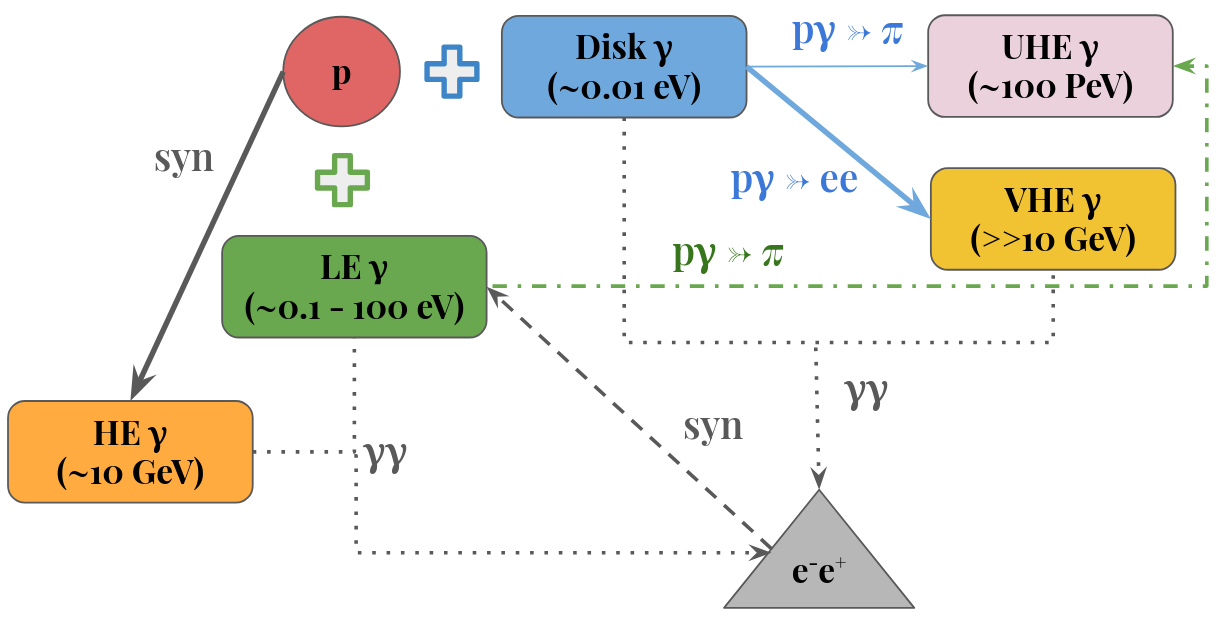}
    \caption{Schematic illustration of interactions among charged particles and photons in a magnetospheric current layer. Solid arrows denote proton synchrotron radiation producing high-energy (HE) photons (orange shaded box), Bethe-Heitler pair production (p$\gamma \rightarrow e^{-}e^{+}$), and photopion production (p$\gamma \rightarrow \pi^{\pm}, \pi^0$). The latter two processes (indicated with blue arrows) rely only on ambient disk photons (blue shaded box), and lead respectively to the generation of very-high-energy (VHE) and ultra-high-energy (UHE) photons (yellow and pink shaded boxes respectively). Dotted lines connect photon populations that interact via photon-photon ($\gamma \gamma$) pair creation, producing secondary pairs (gray shaded triangle). Low-energy (LE) photons (green shaded box) that are emitted by secondary pairs through synchrotron provide an additional target for photopion interactions. Thick arrows indicate the dominant interaction channels.}
    \label{fig:sketch}
\end{figure}

The key interactions among charged particles and photons in the magnetospheric current layer are illustrated schematically in Fig.~\ref{fig:sketch}. Disk photons act as agents for 
Bethe–Heitler pair production, giving rise to energetic radiation. These high-energy photons 
are in turn attenuated by the same disk photon field, triggering an electromagnetic cascade that reshapes the broadband emission.

\begin{table}
\centering
\caption{Baseline model parameters for M87*.}
\label{params}
\begin{tabular}{l l l}
\hline
 Parameters  & Symbol [Units]  &  Values \\
 \hline
\multicolumn{3}{c}{Input} \\
\hline
Black hole mass & $M$ [$M_\odot$] & $10^{9.8}$ \\
Current sheet radius & $r$ [$R_{\rm g}$] & 10 \\
Reconnection rate & $\beta_{\rm rec}$ & 0.1 \\
Fraction of dissipated power into p& $\eta_{\rm p}$ & 0.5 \\
Proton plasma magnetization & $\sigma_{\rm p}$ & $10^{7.5}$ \\
Upstream magnetic field strength & $B$ [G]  & 100 \\
External photon luminosity & $L_{\rm ext}$ [erg/s] & $10^{42}$ \\ 
External peak photon energy & $E_{\rm ext}$ [eV] & $10^{-2}$ \\
Dilution factor of ext. photon field & $f_{\rm d}$ & 1 \\
\hline
\multicolumn{3}{c}{Derived} \\
\hline
Current sheet effective radius & $R$ [cm] & $10^{15.3}$ \\
Dissipated power & $L_{\rm rec}$ [erg/s] &  $10^{44}$ \\
Relativistic proton power  & $L_{\rm p}$ [erg/s] & $10^{43.7}$ \\
Break proton Lorentz factor & $\gamma_{\rm br}$ & $10^7$ \\
Max. proton Lorentz factor & $\gamma_{\rm p, rad}$ & $10^{9.8}$ \\ 
\hline
\end{tabular}
\end{table}

\section{Results}
In this section we present numerical calculations of the broadband photon emission produced in the reconnection layer, highlighting the role of Bethe-Heitler pair production and $\gamma \gamma$ pair production that were not discussed in \href{https://ui.adsabs.harvard.edu/abs/2025ApJ...995L..73H}{H25}. 

Calculations are performed with the leptohadronic radiation code ATHE$\nu$A  \citep{2012A&A...546A.120D}. The code solves a system of coupled partial differential equations describing the evolution of all stable particle distributions inside a spherical, magnetized region, namely pairs, protons, neutrons\footnote{Their decay timescale in the layer rest frame is much longer than their advection timescale, making them effectively a stable species.}, photons, and neutrinos. All interaction rates and emissivities assume isotropic particle distributions. Protons are injected with the same distribution at all times, since acceleration up to the burnoff limit is much faster than $R/c$. We evolve the distributions for a duration equal to the lifetime of the current layer, which is set to $10 \, R/c \simeq 8.5~ R_{15.3} \, \rm d$, motivated by general relativistic MHD simulations \citep{ripperda_black_2022}. The particle escape timescale is $R/c$. The disk photon field is approximated by a gray body with typical photon energy $E_{\rm ext}=0.01$~eV and luminosity $L_{\rm ext}=10^{42}$~erg/s. A more accurate description of its spectrum would not qualitatively affect our results, but we discuss the impact of the disk luminosity in Appendix~\ref{appB}.

\begin{figure}
\centering
\includegraphics[width=0.99\linewidth]{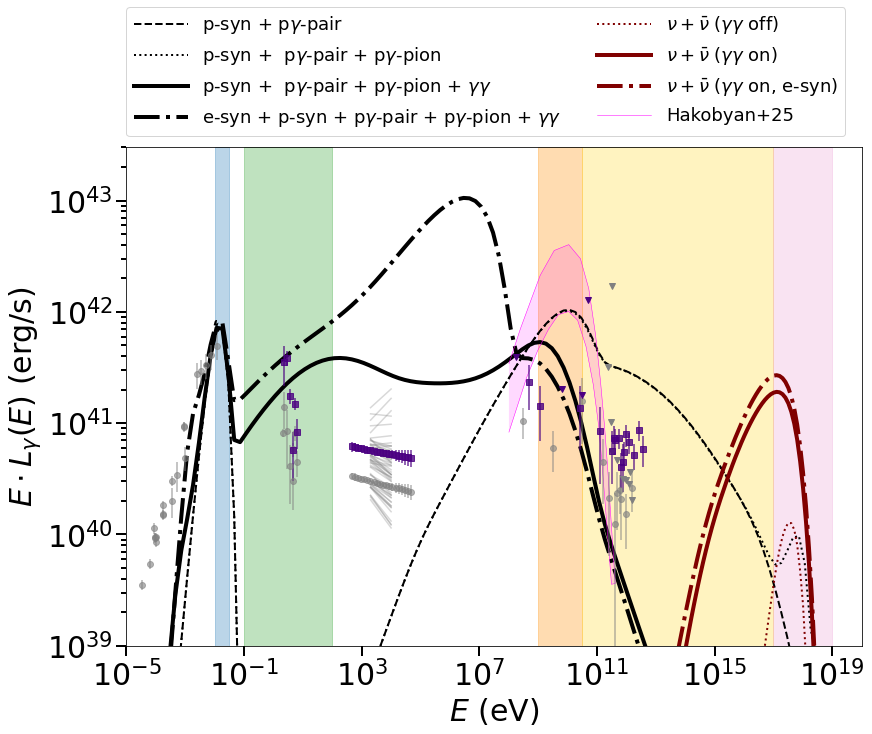}
\caption{Spectral energy distribution of photons (black lines) and neutrinos (dark red lines) produced by relativistic protons accelerated in a magnetospheric current sheet in M87. Thick lines show the escaping photon and neutrino radiation when all processes are taken into account. The magenta-shaded region indicates the range of proton synchrotron spectra presented in \href{https://ui.adsabs.harvard.edu/abs/2025ApJ...995L..73H}{H25}. Energy ranges discussed in text are colored following the coloring scheme in Fig.~\ref{fig:sketch}. Grey and indigo markers show multi-wavelength observations of M87* in 2017 and 2018 at a low and flaring state respectively \citep{2021ApJ...911L..11E, 2024A&A...692A.140A}. Overplotted with grey lines are Chandra X-ray observations from 2007 to 2019 \citep{2023RAA....23f5018C}.}
\label{fig:sed-1}
\end{figure}

Figure~\ref{fig:sed-1} shows the spectral energy distribution (SED) of photons and neutrinos produced in a magnetospheric current layer for the parameter values listed in Table~\ref{params}. Thin lines show spectra computed by selectively including individual physical processes, as indicated in the inset legend, while thick lines correspond to calculations that incorporate all relevant interactions. Emission produced solely from processes involving protons is shown with solid lines, while dash-dotted lines show results when primary pairs\footnote{The pair distribution is modeled as a single power law, $dN_{\rm e}/d\gamma \propto \gamma^{-1}$ for $\gamma \le \gamma_{\rm e, rad}$ where $\gamma_{\rm e, rad} \simeq 10^{6.6} \beta_{\rm rec,-1}^{1/2} B_2^{-1/2}$ is the radiation-reaction limit.} are also taken into account. Attenuation of VHE gamma-rays due to extragalactic background light (EBL) is not included here. The magenta-shaded band presents the range of proton-synchrotron spectra predicted by \href{https://ui.adsabs.harvard.edu/abs/2025ApJ...995L..73H}{H25} \, for comparison. 

We find that synchrotron radiation from Bethe–Heitler pairs produces VHE photons over a broad energy range (yellow shaded region), with a luminosity similar to that estimated in Eq.~\ref{eq:LBH}. Photopion interactions with external photons make a negligible contribution to the photon and neutrino emission (dotted lines), in agreement with the analytical arguments presented in Appendix~\ref{appA2}. In contrast, attenuation of Bethe–Heitler synchrotron photons by disk photons fundamentally alters the picture proposed in \href{https://ui.adsabs.harvard.edu/abs/2025ApJ...995L..73H}{H25}: VHE radiation is efficiently reprocessed to lower energies (green shaded band), attenuating high-energy (HE) photons (orange shaded region) and enhancing neutrino production via photopion interactions (thick dark red line). The resulting photon spectrum (thick solid black line) is characteristic of a synchrotron-dominated electromagnetic cascade \citep{2025JCAP...12..044F}. Including primary pairs produces an additional bright synchrotron component peaking at tens of MeV energies, consistent with \href{https://ui.adsabs.harvard.edu/abs/2025ApJ...995L..73H}{H25}, without modifying the above conclusions (compare thick solid and thick dash-dotted lines). The Bethe–Heitler–initiated cascade is rather insensitive to the detailed spectral shape of the primary pair distribution, since radiation from primary pairs is not a dominant target field for the cascade. In the next section we discuss the apparent tension of the model with multi-wavelength data. \textit{Overall, our results demonstrate that Bethe–Heitler–induced cascades suppress GeV proton-synchrotron emission while significantly enhancing the neutrino output.}

\section{Discussion \& Conclusions}
We presented results for $\sigma_{\rm p} = 3 \times 10^7$, but these are not sensitive to the specific choice. Provided that the proton distribution has a post-break slope $s \lesssim 2.3$ and that the break Lorentz factor of the distribution is lower than the threshold Lorentz factor for Bethe-Heitler interactions with external photons, namely $\gamma_{\rm p,br} = \xi \sigma_{\rm p} < \gamma_{\rm p,th}^{\rm BH} \simeq 10^8 {(x_{\rm ext,-7.7})^{-1}}$ (Eq.~\ref{eq:gp_BH_thr}), the qualitative features of the electromagnetic cascade  -- including the suppression of proton synchrotron emission at tens of GeV energies -- remain unchanged (see Fig.~\ref{fig:seds} in Appendix~\ref{appB}). The neutrino spectrum is likewise robust and continues to peak at energies $\sim \gamma_{\rm p, rad} m_{\rm p} c^2/20 \sim 300$ PeV, largely independent of the precise value of $\sigma_{\rm p}$ as illustrated in the right panel of Fig.~\ref{fig:seds} (see also Fig.~11 of \citealt{karavola_neutrino_2025}). 

The proton distribution injected into our radiative calculations is motivated by PIC simulations of reconnection in pair-dominated plasmas presented in \href{https://ui.adsabs.harvard.edu/abs/2025ApJ...995L..73H}{H25}. Although these simulations did not include Bethe–Heitler pair production, this does not challenge the consistency of the adopted proton distribution, as long as the pair magnetization is much larger than unity and pairs are fast cooling. In the relativistic, pair-dominated reconnection regime relevant here, proton energization is largely decoupled from the pair dynamics. For the parameters considered, the Bethe-Heitler energy-loss timescale remains much longer than both the proton acceleration and escape timescales (see also Eq.~\ref{eq:fBH}), even accounting for cascade photons. Therefore, the proton distribution is not modified by Bethe-Heitler cooling. In other words, the Bethe-Heitler process primarily affects the emergent radiation and pair content of the reconnection layer through electromagnetic cascades, while leaving the proton distribution underlying our calculations unchanged \citep[for similar arguments presented in the context of blazar emission, see][]{2015MNRAS.447...36P}.   

Pair production has previously been explored as a mechanism for plasma enrichment in magnetospheric current layers, through the attenuation of tens of MeV synchrotron radiation emitted by electrons and positrons accelerated during reconnection \citep[e.g.,][]{2023ApJ...944..173C, 2023ApJ...943L..29H, 2024JCAP...12..009S}. 
In the present work, we find that $\gamma\gamma$ attenuation of Bethe–Heitler synchrotron photons provides the dominant channel for pair creation. For the numerical example shown in Fig.~\ref{fig:sed-1}, the primary pair density is $n_{\pm}^{\rm (prim)} \simeq 290~\rm cm^{-3}$, corresponding to an electron magnetization $\sigma_{\rm e}\simeq 10^{6.5} \sim \gamma_{\rm e, rad}$. Including secondary pairs from $\gamma\gamma$ interactions (of hadronic origin) increases the total pair density to $n_{\pm}^{\rm (prim+sec)} \simeq 560 \, n_{\pm}^{\rm (prim)}$, reducing the effective magnetization to $\sigma'_{\rm e}\simeq 10^{3.8} \ll \gamma_{\rm e, rad}$.
In this regime, proton acceleration will not be altered since the plasma remains strongly magnetized and pair dominated (\href{https://ui.adsabs.harvard.edu/abs/2025ApJ...995L..73H}{H25}), but an additional channel of fast particle acceleration becomes relevant in which the pairs
are accelerated while meandering between both sides of the current layer 
\citep{2021ApJ...922..261Z, 2023ApJ...956L..36Z}. The corresponding radiative signatures have been studied in \cite{2024JCAP...12..009S}.  These results demonstrate that Bethe–Heitler–driven electromagnetic cascades can significantly enrich the plasma with pairs and substantially modify the physical conditions in the reconnection layer.

Our results were obtained assuming isotropic particle distributions, a common limitation of radiative codes like ATHE$\nu$A \citep{2026ApJS..282...22C}. This raises concerns about the opacities of interactions that depend on the collision angle of the particles. To address this concern we compare the $\gamma \gamma$ opacity under the isotropic assumption to that of an anisotropic case where only perpendicular collisions are allowed (see Appendix~\ref{app:tau}). We find that the former overestimates the latter opacity by at most $\sim 30\%$ for interactions occurring near the threshold. Since opacities reach $\sim 10^4$ for interactions with disk photons (Eq.~\ref{eq:tau_gg}), this does not affect our main conclusions: the environment remains optically thick to $\gamma$-rays from Bethe-Heitler pairs, whose attenuation triggers the cascade. Nevertheless, a more accurate treatment of anisotropic effects would require (i) a detailed understanding of the proton anisotropy as a function of energy and (ii) a radiative transfer framework that tracks photon geodesics in a realistic current sheet \citep[see e.g.,][]{2025ApJ...985..147S}, both of which are promising directions for future work. 

The results shown in Fig.~\ref{fig:sed-1} are obtained under the assumption of a steady state, corresponding to a current-layer lifetime of $10 R/c \approx 8.5$~d. While this approximation is useful for highlighting the effects of the dominant radiative processes, the formation of the current sheets in the magnetospheric region is inherently time dependent \citep{ripperda_black_2022}. To assess the impact of this assumption, we performed time-dependent calculations (Appendix~\ref{appC}) in which the layer persists for $5 R/c$, while the radiative output is tracked for an additional $5R/c$ as the injection rate of accelerated particles decays exponentially. The resulting time-resolved spectra, shown in Fig.~\ref{fig:sed-2}, reveal an initial enhancement of proton-synchrotron emission at $\sim 10$ GeV, accompanied by synchrotron radiation of Bethe-Heitler pairs up to $\sim 50$ TeV, followed by a delayed brightening at low-energies (0.1-100 eV) and in X rays associated with the reprocessing of VHE photons through electromagnetic cascades. This temporal evolution highlights that the qualitative conclusions drawn from the steady-state treatment remain valid, while predicting spectral delays that may be accessible to multiwavelength observations. Furthermore, these results alleviate the apparent tension between the model and the data shown in Fig.~\ref{fig:sed-1}, since the predicted fluxes evolve on timescales shorter than those typically probed by the archival measurements.

In conclusion, we have shown that even a small fraction of protons in a magnetospheric current layer can significantly alter its observable electromagnetic signatures. Although proton synchrotron radiation can, in principle, produce photons at tens of GeV with luminosities reaching a few percent of the dissipated power, this emission is strongly suppressed by $\gamma\gamma$ pair production. This suppression is enabled by Bethe–Heitler interactions of relativistic protons with disk photons in M87*, which generate very-high-energy photons that initiate electromagnetic cascades. The resulting cascade redistributes the luminosity to lower energies, thereby providing abundant targets for the attenuation of GeV radiation. The same low-energy radiation enhances neutrino production at hundreds of PeV energies. 

The importance of these effects is sensitive to key plasma parameters. They manifest primarily in baryon-poor regions with reconnecting magnetic fields of strength $B \sim 100$~G, weak guide fields (as expected in the equatorial reconnection layer of magnetically arrested disks), and a fraction of dissipated energy transferred to protons $\eta_{\rm p} \gtrsim 0.2$. Outside this regime, proton-synchrotron emission, electromagnetic cascades, and neutrino fluxes are all strongly suppressed. These constraints define a physically motivated parameter space in which Bethe-Heitler pair production in reconnection-driven flares effectively modifies the pure proton-synchrotron GeV emission. Our results highlight the previously underappreciated role of Bethe–Heitler pair production in reconnection-driven flares and motivate future PIC simulations of relativistic reconnection in pair–ion plasmas that explicitly incorporate this process.

\begin{acknowledgments}
The authors thank the anonymous referee for constructive comments that helped clarify several points in the manuscript. The authors would also like to thank Apostolos Mastichiadis, Stamatios I. Stathopoulos, Amir Levinson and Hayk Hakobyan for their useful comments on the manuscript. L.S. acknowledges support from DoE Early Career Award DE-SC0023015 and NASA ATP 80NSSC24K1238. This work was also supported by a grant from the Simons Foundation (MP-SCMPS-00001470) to L.S., and facilitated by Multimessenger Plasma Physics Center (MPPC, NSF PHY-2206609 to L.S.).
\end{acknowledgments}

\begin{contribution}

All authors contributed equally to the manuscript.


\end{contribution}

%



\appendix
\section{Analytical estimates}\label{appA}
In this appendix, we present analytical estimates for the characteristic energies of particles involved in two-particle interactions, including Bethe-Heitler pair production, photopion interactions, and photon-photon ($\gamma \gamma$) pair production. We also derive expressions for the relevant interaction rates to assess the relative importance of each process.

\subsection{Characteristic particle energies}\label{appA1}
In this section, we define the characteristic energies of particles involved in interactions leading to the generation of secondary pairs and subsequently photons across different energies, as illustrated in Fig.~\ref{fig:sketch}. In the following, all photon energies denoted by $x$ are normalized to $m_{\rm e} c^2$. We also introduce the dimensionless magnetic field $b= B/B_{\rm cr}$ where $B_{\rm cr} \equiv m_{\rm e}^2 c^3 / (\hbar e) \simeq 4.4\times 10^{13}$~G.

The typical energy of proton synchrotron photons is $x_{\rm s,p} \approx b \gamma^2 m_{\rm e}/m_{\rm p}$, which translates to $E_{\rm s, p} \simeq 30$~GeV for protons with $\gamma = \gamma_{\rm p, rad}$. The Lorentz factor of protons that satisfy the energy threshold for Bethe-Heitler pair production with the external photons is \citep{1990ApJ...362...38B}
\begin{equation}
\gamma^{\rm BH}_{\rm p, th} = \frac{2}{x_{\rm ext}} = 10^8 {(x_{\rm ext, -7.7})^{-1}},
\label{eq:gp_BH_thr}
\end{equation}
which lies above the break Lorentz factor of the distribution if $\sigma_{\rm p} < 10^8 \xi^{-1} {(x_{\rm ext, -7.7})^{-1}}$ and below the proton synchrotron burnoff limit. 

The pairs produced from these on-threshold interactions have a typical Lorentz factor $\gamma^{\rm BH}_{\rm e} \approx \gamma^{\rm BH}_{\rm p, th}$ \citep{2024JCAP...07..006K}. This is typically higher than the maximum Lorentz factor reached by the primary pairs accelerated in the reconnection layer, $\gamma_{\rm e, rad} = (6 \pi e \beta_{\rm rec}/ (\sigma_{\rm T} B))^{1/2} \approx 10^{6.6} \, {(\beta_{\rm rec, -1})^{1/2} (B_{2})^{-1/2}}$. As a result, Bethe-Heitler pairs emit synchrotron photons, $x_{\rm s, e}^{\rm BH} = b (\gamma_{\rm e}^{\rm BH})^2$,  well beyond the synchrotron burn-off limit of accelerated pairs in the layer, 
\begin{equation}
  x_{\rm s, e}^{\rm BH} = \frac{4 b}{x_{\rm ext}^2}  \simeq  10^{4.3} B_2 \, {(x_{\rm ext, -7.7})^{-2}} \Rightarrow E_{\rm s, e}^{\rm BH}  \simeq  10 \, B_2 \, {(x_{\rm ext, -7.7})^{-2}} \, \rm GeV.
  \label{eq:xs_BH}
\end{equation}

The interactions between protons with $\gamma > \gamma^{\rm BH}_{\rm p, th}$  and external photons occur far above the energy threshold, resulting in the production of extremely energetic pairs \citep[see Fig.~3 in][]{2024JCAP...07..006K}. The highest-energy Bethe-Heitler pairs are therefore expected to reach Lorentz factors $\gamma^{\rm BH}_{\rm e,\max} \sim 0.1\, \gamma_{\rm p, rad} (m_{\rm p}/m_{\rm e}) \simeq 10^{12} \, {(B_2)^{1/2}}$~\citep{2008PhRvD..78c4013K}, where the numerical prefactor 0.1 is empirically determined \citep[see e.g., Fig.~3 of][]{2024JCAP...07..006K}. Therefore, synchrotron emission from Bethe-Heitler pairs spans a broad energy range, from 10~GeV to 0.5~EeV ($\approx \gamma^{\rm BH}_{\rm e, \max} m_{\rm e} c^2$). While the detailed shape of the resulting synchrotron spectrum is difficult to predict analytically, due to its dependence on both the proton and target photon distributions \citep[see e.g. Fig.~4 in][]{2024JCAP...07..006K}, it is robust to conclude that synchrotron radiation of Bethe-Heitler pairs dominates over inverse Compton emission from primary pairs in the current layer at TeV energies, as instead proposed by \href{https://ui.adsabs.harvard.edu/abs/2025ApJ...995L..73H}{H25}. The reason for this dominance is twofold: Bethe-Heitler pairs are more energetic than primary pairs, hence radiate more efficiently, and the magnetic energy density greatly exceeds that of the external radiation field ($u_{\rm B} \gg u_{\rm ext}$).

Energetic photons are susceptible to attenuation through interactions with low-energy photons, leading to production of secondary pairs, as illustrated in Fig.~\ref{fig:sketch}.  We therefore determine the energy of $\gamma$-ray photons that pair-produce on threshold with the external photons 
\begin{equation}
    x_{\gamma, \rm th} = \frac{2}{x_{\rm ext}} = 10^8 {(x_{\rm ext, -7.7})^{-1}} \Rightarrow  E_{\gamma, \rm th} = 50 \, {(x_{\rm ext, -7.7})^{-1}} \, \rm TeV.
    \label{eq:x_gg_th}
\end{equation}
Photons near this energy experience the strongest attenuation, as they interact with target (disk) photons of the highest number density and because the $\gamma \gamma$ cross section peaks just above threshold \citep{1990MNRAS.245..453C}. Bethe-Heitler synchrotron photons at lower energies (see Eq.~\ref{eq:xs_BH}) may also be attenuated through interactions with target photons of energy $x > 2/x_{\rm s, e}^{\rm BH} = 10^{-4}$, corresponding to $E > 50$~eV. Such optical-ultraviolet photons can originate either from the accretion disk or be produced locally in the reconnection layer by primary pairs. However, this attenuation is expected to be weaker than that affecting photons with energies near $x_{\gamma,\rm th}$, owing to the much lower number density of target photons above $\sim 50$ eV.  We provide some estimates for the $\gamma \gamma$ optical depth in the next section. 

Attenuation of energetic photons leads to the production of secondary pairs. The typical Lorentz factor of pairs produced by $\gamma \gamma$ interactions close to the threshold is $\gamma_{\rm e}^{\gamma \gamma} \approx  x_\gamma /2$
\begin{equation}
\gamma_{\rm e}^{\gamma \gamma} = \left \{ 
\begin{tabular}{cc}
$x_{\rm ext}^{-1} = 5\times 10^7 \, {(x_{\rm ext, -7.7})^{-1}}$,     & $x_\gamma = x_{\gamma, \rm th}$ \\ \\
$2b \, x_{\rm ext}^{-2} \simeq 10^4 \, B_2 {(x_{\rm ext, -7.7})^{-2}}$,    & $x_\gamma = x_{\rm s, e}^{\rm BH}$
\end{tabular}
\right.
\end{equation}

Synchrotron photons emitted by pairs produced through $\gamma \gamma$ interactions have typical energies $x_{\rm s, e}^{\gamma \gamma} =b (\gamma_{\rm e}^{\gamma \gamma})^2$, which correspond to 
\begin{equation}
E_{\rm s, e}^{\gamma \gamma}  \simeq \left \{
\begin{tabular}{c c}
 $3 \, B_2 {(x_{\rm ext,-7.7})^{-2}}\rm GeV$,    & $\gamma_{\rm e}^{\gamma \gamma} = x_{\gamma, \rm th}/2$ \\ \\
$150~B_2^3 {(x_{\rm ext, -7.7})^{-4}}~\rm eV$,   & $\gamma_{\rm e}^{\gamma \gamma} = x^{\rm BH}_{\rm s,e}/2$
\end{tabular}
\right. 
\end{equation}
Moreover, synchrotron cooling drives secondary pairs to Lorentz factors well below their injection values, leading to the production of photons with energies $\lesssim 150$ eV. As a result, pairs generated through $\gamma\gamma$ interactions between Bethe–Heitler synchrotron photons and external photons act as an effective mediator for enhancing the low-energy photon field ($\sim 0.1-150$~eV), thus indirectly increasing the photopion production rate (see also Fig.~\ref{fig:sketch}).

The energy threshold condition for photopion production reads $\gamma_{\rm p}  x \ge  m_{\pi}/m_{\rm e} \simeq 280$ \citep{1990ApJ...362...38B}. The photon energy that interacts at threshold with the most energetic protons in the distribution is then
\begin{eqnarray}
x^{p\gamma}_{\rm th} &=& \frac{280}{\gamma_{\rm p, rad}} = 4 \times 10^{-8} \, {(\beta_{\rm rec,-1})^{-1/2} (B_2)^{1/2}} \Rightarrow \nonumber \\ 
E^{p\gamma}_{\rm th} &=& 0.02~{(\beta_{\rm rec, -1})^{-1/2} (B_2)^{1/2}} \, \rm eV
\label{eq:x_pg_th}
\end{eqnarray}
which is just above the characteristic energy of disk photons. Beyond the $\Delta^+$ resonance, the effective photopion cross section remains approximately constant \citep{1990ApJ...362...38B}. As a result, protons with Lorentz factor $\gamma_{\rm p, rad}$ will interact with comparable probability with more energetic photons. Therefore, an increase in the number density of photons with energies $>0.02$~eV leads to a corresponding enhancement of the photopion production rate. We present estimates of the photopion efficiency in the next section. 

\subsection{Interaction rates}\label{appA2}
In this section, we provide estimates for the interaction rates of Bethe-Heitler and photopion interactions as well as for the $\gamma \gamma$ optical depth when the external (disk) photons are the dominant targets. 

The fraction of energy transferred from protons to Bethe-Heitler pairs, $f_{\rm BH}$, can be estimated as 
\begin{eqnarray}
\label{eq:fBH}
f_{\rm BH} &=& \min(1, t^{-1}_{\rm BH} t_{\rm dyn}) = \min\left(1, \hat{\sigma}_{\rm BH} R n_{\rm ext} \right) \\ \nonumber
& \simeq& 0.06 \, f_{\rm d, 0} \, L_{\rm ext, 42} \, {(R_{15.3})^{-1}} \,  x_{\rm ext, -7.7} 
\end{eqnarray}
where $t_{\rm dyn} = R/c$,  $n_{\rm ext} \approx u_{\rm ext}/E_{\rm ext}$ is the number density of external photons, and $\hat{\sigma}_{\rm BH} \simeq 8\times 10^{-31}~\rm cm^2$  is the maximum effective cross section accounting for the  inelasticity of the interaction \citep{1992ApJ...400..181C}. Therefore, about 6\% of the luminosity carried by protons with Lorentz factor $\gamma_{\rm p, \rm th}^{\rm BH}$  (these are interacting on threshold with the external photons) is transferred to Bethe-Heitler pairs 
\begin{equation}
L_{\pm}^{\rm BH} \approx f_{\rm BH} \left[\gamma L_{\rm p}(\gamma) \right]_{\gamma=\gamma^{\rm BH}_{\rm p, \rm th}} = \frac{ f_{\rm BH}  L_{\rm p}}{1 + \ln \left(\frac{\gamma_{\rm p,\max}}{\gamma_{\rm p, br}} \right)} \simeq 4.4\times 10^{41}~\rm erg/s
\label{eq:LBH}
\end{equation}
This estimate does not account for far-from-threshold interactions, which also contribute to more energetic Bethe-Heitler pairs as described in the previous section. As a result, Eq.~\ref{eq:LBH} should be considered a lower limit.

The most energetic protons in the distribution interact at the photopion threshold with photons $\sim 0.02$~eV (see Eq.~\ref{eq:x_pg_th}). The photopion cross section peaks at the $\Delta^+$ resonance, at interaction energies approximately three times higher than the threshold \citep{1990ApJ...362...38B}, implying that the most probable interactions would require photons with energies $>$0.06 eV,  well above the characteristic energy of the disk photon field considered here. Consequently, photopion interactions with disk photons do not constitute an efficient energy-loss channel for protons, in agreement with the estimates of \href{https://ui.adsabs.harvard.edu/abs/2025ApJ...995L..73H}{H25}.

Finally, the optical depth for the attenuation of $\gamma$-ray photons interacting near threshold with external photons is
\begin{eqnarray}
\label{eq:tau_gg}
\tau_{\gamma \gamma} \approx \sigma_{\gamma \gamma} R  n_{\rm ext} \simeq
5 \times 10^4 f_{\rm d, 0} \, L_{\rm ext, 42} \, {(R_{15.3})^{-1}} \,  x_{\rm ext, -7.7} 
\end{eqnarray}
where we adopt the peak value of the $\gamma\gamma$ cross section appropriate for interactions just above threshold \citep{1990MNRAS.245..453C}. Therefore, the emission at $E_{\gamma, \rm th} = 50$~TeV  (see Eq.~\ref{eq:x_gg_th}) will be strongly suppressed by a factor of $\tau_{\gamma \gamma}^{-1}$. The absorbed luminosity, $(1 - 1/\tau_{\gamma \gamma}) \, [EL_{\gamma}(E)]_{E=E_{\gamma, \rm th}}$, is instead reprocessed through the injection of secondary pairs and subsequently radiated at lower energies, as illustrated in Fig.~\ref{fig:sketch}. 

In the adopted spherical geometry all optical depths are defined with respect to the radius of the sphere (see, for example, Eq.~\ref{eq:tau_gg}),  where  $R \sim r/5$ based on the equivalent volume mapping introduced in Sec.~\ref{sec:model}.  In cylindrical geometry, the relevant path length instead depends on direction, ranging from $\sim H/2 \simeq 0.25 R$ for photons escaping vertically to $\sim r/2 \simeq 2.5 R$ for photons propagating radially, along the layer. The optical depth obtained in the spherical approximation therefore lies between the minimum and maximum values expected in a cylindrical geometry. As a result, the spherical approximation neither systematically overestimates nor underestimates $\gamma \gamma$ absorption, but instead provides an angle-averaged opacity that brackets the true anisotropic values. A fully anisotropic radiative transfer treatment would require a non-spherical kinetic solver and is beyond the scope of the present work.

\section{Effects of model parameters}\label{appB}
In this section, we demonstrate the impact of key model parameters on the photon spectral energy distribution (SED) and the neutrino emission. We vary the magnetic field strength $B$ in the upstream region of the magnetospheric current layer, the disk luminosity $L_{\rm ext}$, the post-break slope of the proton distribution $s$, the proton plasma magnetization $\sigma_{\rm p}$,  and the fraction of energy received by protons $\eta_{\rm p}$ (one parameter at a time).

For the baseline model, we adopted $B=100$~G, if M87* is in a magnetically arrested state~\citep[e.g.,][]{2021MNRAS.507.4864Y, 2023ApJ...944..173C, 2024JCAP...12..009S, 2025ApJ...995L..73H}. In this regime, transient magnetospheric current layers may be formed \citep{ripperda_black_2022}. However, modeling of the total and polarized emission from M87* (within an one-zone model of relativistically hot electron plasma) yields $B\lesssim 30$~G \citep{2021ApJ...910L..13E}. Therefore, we performed numerical calculations with $B=30$~G and $B=10$~G, while keeping all other input parameters fixed to their baseline values (Table~\ref{params}).

\begin{figure}
    \centering
    \includegraphics[width=0.99\linewidth]{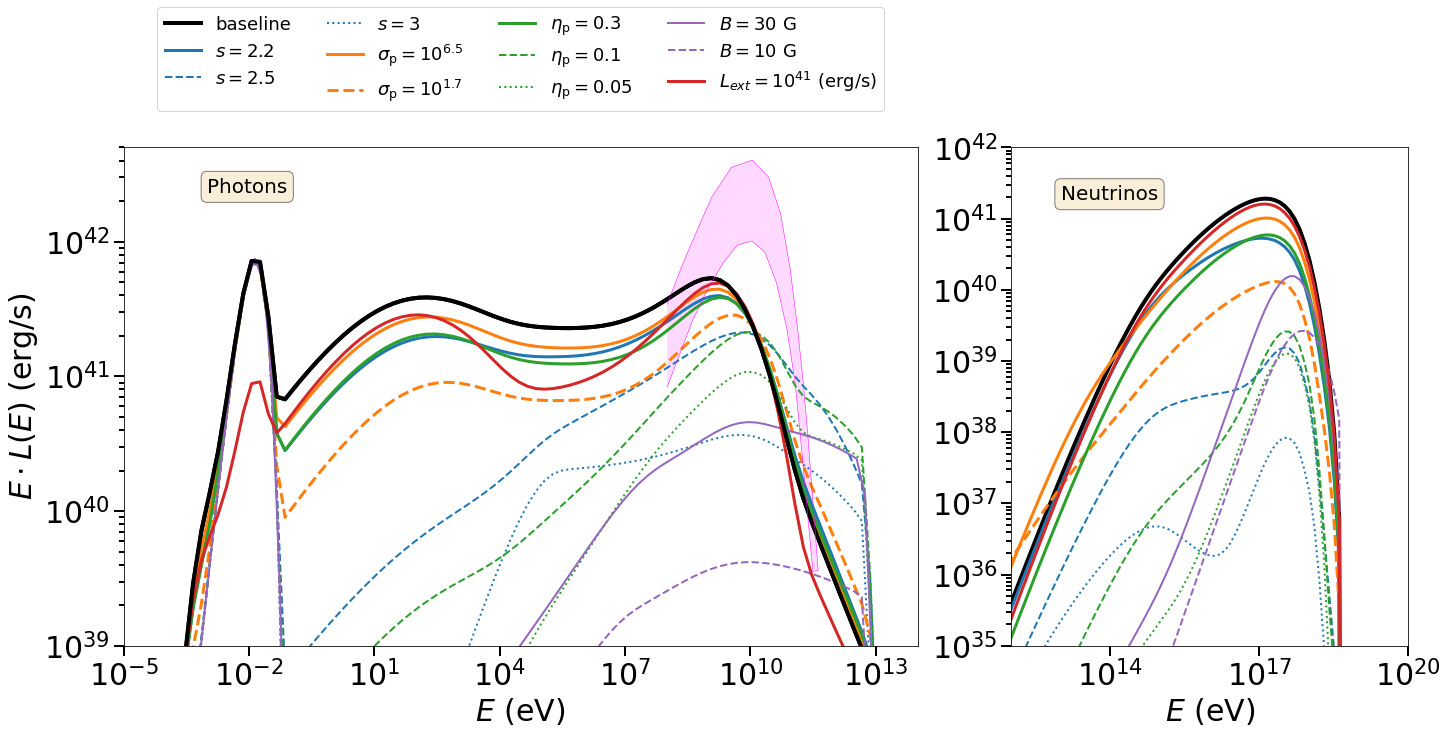}
    \caption{Spectral energy distribution of photons (left) and neutrinos of all flavors (right) produced by relativistic protons accelerated in a magnetospheric current sheet in M87*. Each model is calculated for the same parameters as those of the baseline model except for one parameter that is varied, as shown in the inset legend. Models that differ significantly from the baseline model are plotted with thin lines.}
    \label{fig:seds}
\end{figure}

The results are presented in Fig.~\ref{fig:seds} with purple solid and dashed lines. The SED of the emerging radiation differs substantially from that of the baseline model. The power injected into relativistic protons is reduced, since $L_{\rm p}\propto B^2$, leading to weaker proton synchrotron emission. Although the opacity for Bethe-Heitler pair-production on the external photons is unaffected (see Eq.~\ref{eq:fBH}), the luminosity transferred to pairs is lower (see Eq.~\ref{eq:LBH}). As a result, the absorbed VHE luminosity available for reprocessing to lower energies is also reduced. Because the synchrotron cooling rate of secondary pairs produced in $\gamma \gamma$ interactions decreases, the resulting number density of cascade photons between 0.1 - 100 eV is many orders of magnitude lower than in the baseline model. Therefore, the GeV emission is not attenuated and the neutrino emission is not enhanced, in contrast to the baseline model. However, the luminosity at tens of GeV energies is not high enough to explain the observed GeV flaring emission, as proposed by \href{https://ui.adsabs.harvard.edu/abs/2025ApJ...995L..73H}{H25}.

Next we tested the role of external photon density. To this end, we reduced by an order of magnitude the luminosity of the disk photon field (or equivalently, used a dilution factor of 10\% while keeping $L_{\rm ext}$ fixed to its baseline value). This value may be representative of the disk radiation during flux eruption episodes in which the disk is truncated at larger radii during the formation of the current layer. The results are shown in Fig.~\ref{fig:seds} with solid red lines. The neutrino emission and the photon emission above 1~GeV are nearly identical to the baseline model. However, the shape of the cascade spectrum differs from the baseline case in that the two synchrotron components, one from the secondary pairs at low energies (peaking at approximately 100 eV) and the other from protons at GeV energies, are clearly separated. In particular, the secondary synchrotron spectrum shows
a cutoff at $\sim 0.3$ keV. Therefore, X-ray observations from keV to tens of keV energies have the potential to constrain the density of external photons. 

The post-break proton slope depends on the strength of the non-reconnecting magnetic field component, known as the guide field. Steeper post-break particle spectra are found for stronger guide fields \citep[e.g.][]{2017ApJ...843L..27W, 2023ApJ...959..137C, 2024ApJ...972....9C}. We therefore tested the sensitivity of our results to the post-break slope of the proton distribution by adopting $s=2.2, 2.5$ and 3. For $s=2.2$ the results are qualitatively the same as in the baseline model except for the lower luminosity of the cascade and the neutrino emission. This reduction reflects the lower power carried by protons with $\gamma_{\rm p}\ge \gamma_{\rm p, th}^{\rm BH} > \gamma_{\rm p, br}$. In contrast, the cascade and neutrino emissions are strongly suppressed if $s\ge 2.5$. In conclusion, the Bethe-Heitler initiated cascade may be of relevance in pair-dominated reconnection in the weak-guide field regime. The reconnection layer formed during flux eruption events is indeed expected to have a vanishing guide field.

\begin{figure}
\includegraphics[width=0.99\linewidth]{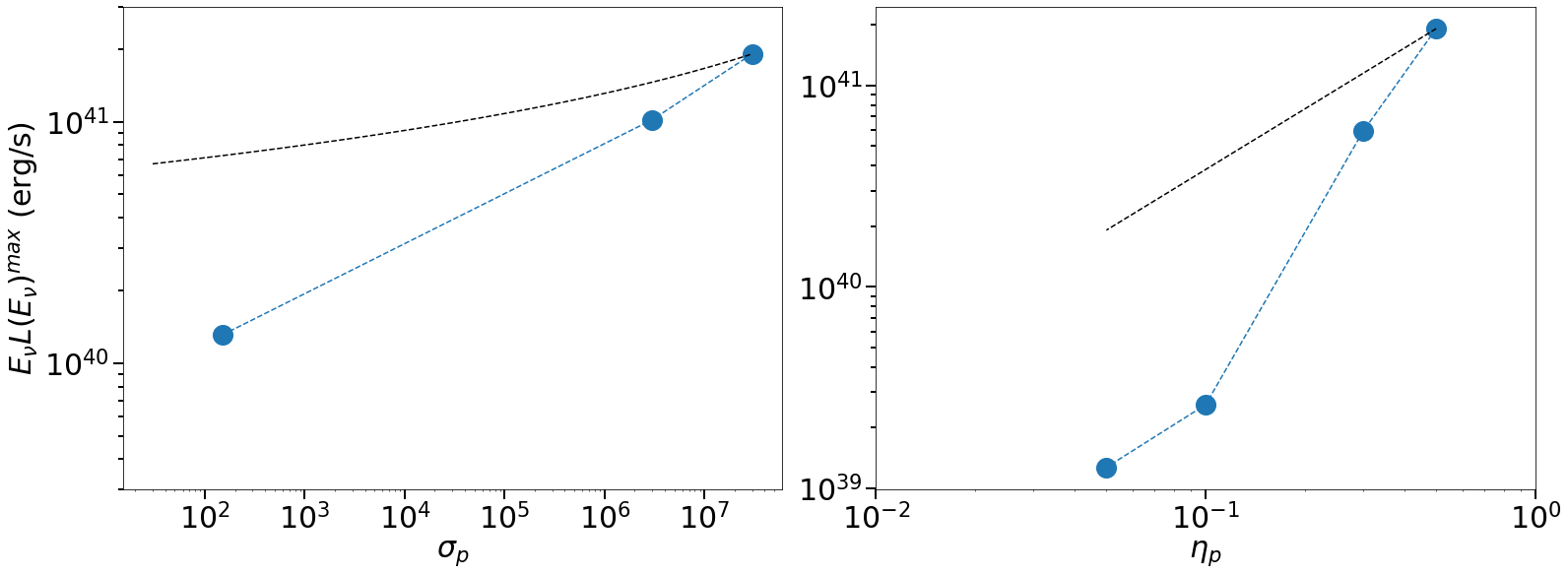}
\caption{Peak neutrino luminosity plotted against $\sigma_{\rm p}$ (left panel) and $\eta_{\rm p}$ (right panel). Dashed black lines show simple analytical scalings that do not account for non-linear effects.}
\label{fig:scalings}
\end{figure}

The high proton plasma magnetization used in the baseline model relies on the presence of a baryon-poor cavity above and below the equatorial current layer. The physical plasma density in these regions is difficult to measure because of the artificial density floors applied to general relativistic magnetohydrodynamic (GRMHD) simulations \citep[e.g.][]{ripperda_black_2022}. Axisymmetric GRMHD studies tracking baryon transport from the disk to the magnetosphere using a passive tracer suggest that extended high-$\sigma_{\rm p}$ regions can form (Chow et al., priv. comm.). Nevertheless, it is instructive to examine the cascade for lower proton magnetization values, which may be relevant if evacuation of disk plasma from the magnetospheric current layers is incomplete. We choose $\sigma_{\rm p}= 10^{6.5}$ and $50$ (orange solid and dashed lines). The lower magnetization case corresponds to an upstream region filled with plasma of proton density $n_{\rm p} = 10^{4}~\rm cm^{-3}$, representative of the disk in M87* \citep[e.g.,][]{2009ApJ...697.1164B, 2019ApJ...875L...5E}. For $\sigma_{\rm p} = 10^{6.5}$, the proton synchrotron luminosity remains essentially unchanged, as it is dominated by protons accelerated to the synchrotron burnoff limit. Instead, the luminosity of the electromagnetic cascade is moderately reduced, since it is driven by Bethe–Heitler interactions of protons with $\gamma > \gamma_{\rm p, th}^{\rm BH}$ interacting with disk photons. Lowering $\sigma_{\rm p}$ while keeping $L_{\rm p}$ fixed reduces the fraction of the injected proton power available for Bethe–Heitler interactions, thereby diminishing the cascade luminosity.
Adopting an even lower proton magnetization, $\sigma_{\rm p} = 50$, yields qualitatively similar results to the higher-$\sigma_{\rm p}$ cases, although the overall emission luminosity is reduced by a factor of $\sim10$ (for neutrinos, in particular, see left panel in Fig.~\ref{fig:scalings}).

For the baseline model, we considered an optimistic value for the fraction of dissipated energy transferred to protons, $\eta_{\rm p}=0.5$, in order to explore additional physical processes in regimes where proton-synchrotron emission is significant. More realistic values for reconnection in pair-dominated plasmas relevant to our work are $\eta_{\rm p} \sim 0.1-0.2$~\citep{2019ApJ...880...37P, 2025ApJ...995L..73H}. To illustrate the effect of varying $\eta_{\rm p}$, we also consider $\eta_{\rm p} = 0.3$, $0.1$, and $0.05$, with the corresponding results shown by the green lines in Fig.~\ref{fig:seds}. For $\eta_{\rm p} \le 0.1$, the proton synchrotron luminosity decreases, and both the cascade and neutrino emission are suppressed, as the protons receive less total power. The dependence of the neutrino emission on $\eta_{\rm p}$ is not strictly linear, because neutrino production is partially enhanced by photons produced in the cascade.

Finally, Fig.~\ref{fig:scalings} demonstrates the non-linearity of  neutrino production on $\sigma_{\rm p}$ and $\eta_{\rm p}$. In the left panel, the dashed line indicates the expected reduction in peak neutrino luminosity due to the decreased power carried by protons at the synchrotron burnoff limit, which scales as $(1 + \ln(\gamma_{\rm p, rad}/\gamma_{\rm p, br}))^{-1}$. For $\sigma_{\rm p}=50$, the peak neutrino luminosity is suppressed by roughly two orders of magnitude more than this simple estimate, reflecting its additional dependence on the cascade photon density. Similarly, the right panel shows the dependence on $\eta_{\rm p}$ where the black dashed line has a slope of unity, corresponding to a linear scaling in the absence of cascade effects. Once again, the dependence found in our calculations is steeper.

\section{Photon-photon pair production opacity for anisotropic collisions}\label{app:tau}
In this section we compare the opacity for $\gamma \gamma$ pair production under the assumption of isotropic collisions and nearly perpendicular collisions.

The optical depth per unit length for a $\gamma$-ray photon of energy $E_\gamma$ is defined as:
\begin{equation}
\frac{d\tau_{\gamma \gamma} (E_\gamma)}{d\ell} = \int d\epsilon \int d\Omega (1-\cos \theta) \, n(\epsilon, \Omega) \sigma_{\gamma \gamma}(S)
\label{eq:tgg-general}
\end{equation}
where $n(\epsilon, \Omega) = \frac{dn}{d\Omega d\epsilon}$ is the differential number density of target photons, and $S = E_\gamma \epsilon (1-\cos \theta)/(m_e c^2)^2$ is the interaction energy. For analytical calculations, we will use the approximate expression for the cross section, $\sigma_{\gamma \gamma} \approx \sigma_0 H(S-2)/S$, where $H(x)$ is the Heaviside function. This approximation is reliable for $S \gg 2$~\citep{2009HEA_BH}.

\textit{Isotropic collisions with monoenergetic target photon field.} The differential density is written as $n_{\rm iso}(\epsilon, \Omega) = \frac{n_0}{4\pi} \delta(\epsilon- \epsilon_0)$. Substitution in Eq.~\ref{eq:tgg-general} results in 
    
    \begin{equation}
    \frac{d\tau^{(iso)}_{\gamma \gamma} (E_\gamma)}{d\ell} \approx n_0 \sigma_0 \frac{(m_e c^2)^2}{E_\gamma \epsilon} \left(1 -\frac{(m_e c^2)^2}{E_\gamma \epsilon}  \right)  H\left(E_\gamma \epsilon_0 -  (m_e c^2)^2  \right).
    \end{equation}

\textit{Perpendicular collisions with monoenergetic target photon field.} Here, the disk emits isotropically, but we are conditioning on photons that intersect the current sheet (on the equatorial plane) at an angle $\pi/2$. The differential density is written as $n(\epsilon, \Omega) = n_0\delta(\Omega  - \Omega_{\pi/2}) \delta(\epsilon- \epsilon_0)$. Substitution in Eq.~\ref{eq:tgg-general} results in 
    \begin{equation}
    \frac{d\tau^{(\pi/2)}_{\gamma \gamma} (E_\gamma)}{d\ell} \approx n_0 \sigma_0 \frac{(m_e c^2)^2}{E_\gamma \epsilon_0} H\left(E_\gamma \epsilon_0 - 2(m_e c^2)^2  \right)
    \end{equation}
    Let $x = E_\gamma \epsilon_0 / (m_e c^2)^2$. By comparing the two expressions for the optical depths, we find 
    \begin{equation}
        \frac{d\tau^{(iso)}_{\gamma \gamma}}{d\tau^{(\pi/2)}_{\gamma \gamma}} \approx 1-1/x \approx 1, x \gg 1.
    \end{equation}
    Near the threshold ($x \gtrsim 2$) we find that $d\tau^{(\pi/2)}_{\gamma \gamma}/d\ell \approx 2 \, d\tau^{(iso)}_{\gamma \gamma}/d\ell$, but at these interaction energies our approximation for the cross section becomes less accurate. 
     
    \begin{figure}
        \centering
        \includegraphics[width=0.55\linewidth]{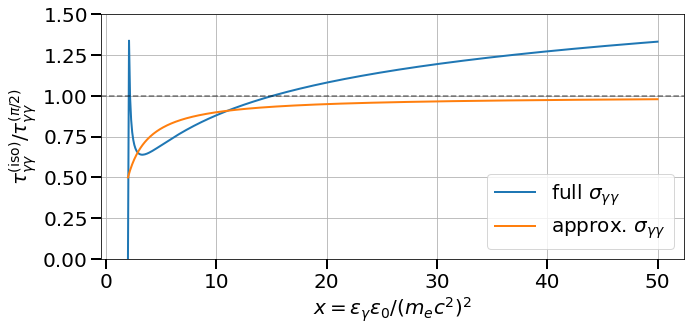}
        \caption{Ratio of optical depths for isotropic and perpendicular only collisions.}
        \label{fig:tgg}
    \end{figure}
    
    If we numerically integrate Eq.~\ref{eq:tgg-general} using the full expression of the $\gamma \gamma $ cross section~\citep{2009HEA_BH}, we find that for $x \gtrsim 2$ the assumed isotropy in collisions overestimates the opacity for perpendicular collisions by a factor of $30\%$ at most.  Given that the opacities we find assuming isotropic collisions are $\sim 10^4$ (see e.g. Eq. A9), the error associated with this assumption is small and will not alter our main conclusions about redistribution of photon energy through non-linear cascades.

\section{Time-dependent calculations}\label{appC}
In this section, we present time-dependent calculations of a reconnection-driven flare for the baseline model; its parameters are listed in Fig.~\ref{fig:sed-1}. To mimic a transient current sheet with lifetime $T$, we assume that the power injected into relativistic protons remains constant up to $T$ and then decreases exponentially, i.e. $L_{\rm p}(t) \propto e^{-c(t-T)/R}$ for $t \ge T$. Protons are injected with the same distribution, as the acceleration up to the burnoff limit is much faster than the dynamical timescale $R/c$. 

\begin{figure}
\centering
\includegraphics[width=0.47\linewidth]{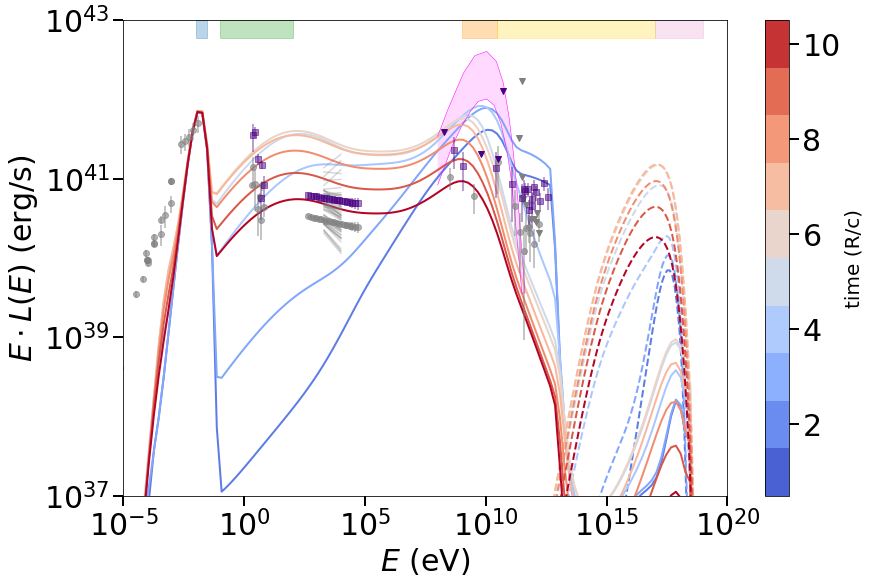}
\includegraphics[width=0.47\linewidth]{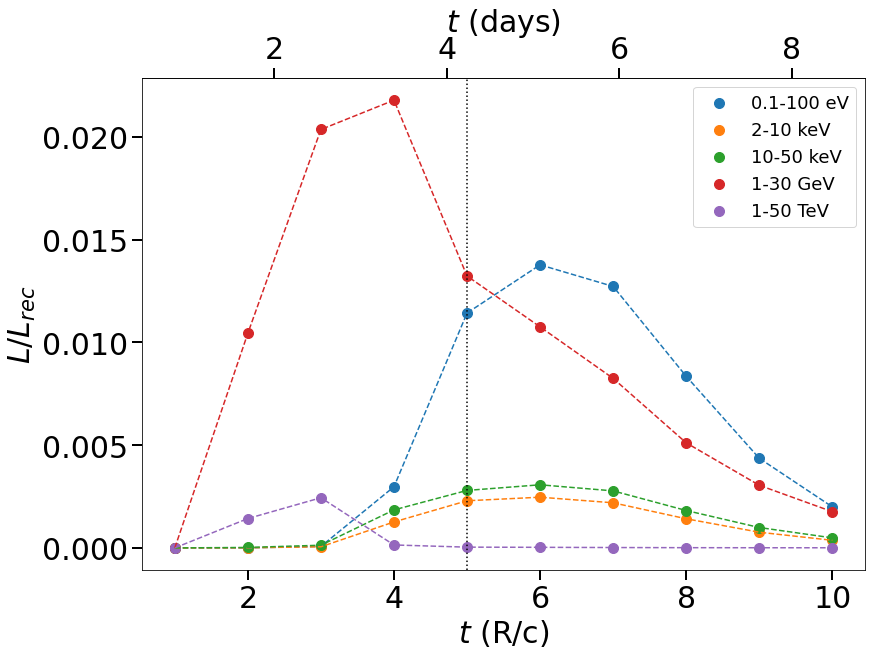}
\caption{Time-dependent emission from a reconnection-driven flare in M87*. \textit{Left panel:} Snapshots of the spectral energy distribution of photons (solid lines) and neutrinos (dashed lines) produced by relativistic protons accelerated in a magnetospheric current sheet in M87. The magenta-shaded region indicates the range of proton synchrotron spectra presented in \href{https://ui.adsabs.harvard.edu/abs/2025ApJ...995L..73H}{H25}. Energy ranges discussed in text are colored following the coloring scheme in Fig.~\ref{fig:sketch}. Grey and indigo markers show multi-wavelength observations of M87* in 2017 and 2018 during a low and flaring state respectively \citep{2021ApJ...911L..11E, 2024A&A...692A.140A}. Chandra X-ray observations from 2007 to 2019 are shown with grey lines \citep{2023RAA....23f5018C}. \textit{Right panel:} Light curves of a reconnection-driven flare in M87*. Luminosities are shown in four representative energy bands (see inset legend) and are normalized to the power dissipated in the current layer. Colors are not intended to correspond to those used in the left panel.}
\label{fig:sed-2}
\end{figure}

Figure~\ref{fig:sed-2} shows the temporal evolution of the photon and neutrino spectral energy distributions (left panel), as well as representative light curves, for $T = 5R/c$ (right panel). We find that proton synchrotron emission in the 1–30~GeV band reaches its maximum at $t \sim 4R/c$, with $L_{1-30~\rm GeV} \simeq 0.02 \, L_{\rm rec}$, while TeV emission -- driven by Bethe–Heitler pair synchrotron radiation -- is strongly suppressed due to $\gamma\gamma$ attenuation. In contrast, the very-high-energy radiation that is reprocessed into lower energies peaks at later times (see the orange and blue curves in the right panel). This late-time enhancement of the 0.1–100~eV photon density leads to a fast decline of the GeV emission and a corresponding increase in the neutrino luminosity (not shown in the light-curve plot, but inferred from the SED evolution in the left panel).

\bibliography{bibliography}{}
\bibliographystyle{aasjournalv7}



\end{document}